\documentclass[8.5pt]{article}
\usepackage{spconf,amsmath,graphicx}
\usepackage{multirow}
\usepackage{breqn}
\usepackage{amssymb}
\usepackage{adjustbox}
\usepackage{booktabs}
\usepackage{makecell}
\usepackage{listings}
\usepackage{xcolor}
\usepackage{url}

\definecolor{codegreen}{rgb}{0,0.6,0}
\definecolor{codegray}{rgb}{0.5,0.5,0.5}
\definecolor{codepurple}{rgb}{0.58,0,0.82}
\definecolor{backcolour}{rgb}{1, 1, 1}

\lstdefinestyle{mystyle}{
    backgroundcolor=\color{backcolour},   
    commentstyle=\color{codegreen},
    keywordstyle=\color{magenta},
    numberstyle=\tiny\color{codegray},
    stringstyle=\color{codepurple},
    basicstyle=\ttfamily\footnotesize,
    breakatwhitespace=false,         
    breaklines=true,                 
    captionpos=b,                    
    keepspaces=true,                 
    numbers=left,                    
    numbersep=5pt,                  
    showspaces=false,                
    showstringspaces=false,
    showtabs=false,                  
    tabsize=2
}
\title{ConVoiFilter: A CASE STUDY OF DOING COCKTAIL PARTY SPEECH RECOGNITION}

\name{Thai-Binh Nguyen$^{1}$, Alexander Waibel$^{1,2}$}

\address{$^{1}$Karlsruhe Institute of Technology\\$^{2}$Carnegie Mellon University\\
\texttt{thai-binh.nguyen@kit.edu}}
\begin{document}
%
\maketitle
\begin{abstract}


This paper presents an end-to-end model designed to improve automatic speech recognition (ASR) for a particular speaker in a crowded, noisy environment. The model utilizes a single-channel speech enhancement module that isolates the speaker’s voice from background noise (ConVoiFilter) and an ASR module. The model can decrease ASR's word error rate (WER) from 80\% to 26.4\% through this approach. Typically, these two components are adjusted independently due to variations in data requirements. However, speech enhancement can create anomalies that decrease ASR efficiency. By implementing a joint fine-tuning strategy, the model can reduce the WER from 26.4\% in separate tuning to 14.5\% in joint tuning. We openly share our pre-trained model to foster further research \url{hf.co/nguyenvulebinh/voice-filter}.

\end{abstract}
\begin{keywords}
ASR, Speech Enhancement, Voice Filter
\end{keywords}
\section{Introduction}
\vspace{-2mm}

In the ideal environment with a close speaking microphone, the speech recognition performance of current ASR systems can surpass humans, with a common word error rate usually below 5\% \cite{DBLP:conf/interspeech/NguyenSW21}. However, in a realistic environment, the ASR system has to deal with complex acoustic conditions like noise, reverberation, cross-talk (known as the cocktail party setting) \cite{yang1999multimodal, bett2000multimodal}. As a result, it makes the model performance drops dramatically. For example, in CHiME-5 competition, WER of barely below 80\% achieved by the baseline system; using a robust back-end, approximately 60\% WER is achieved \cite{zorilua2019investigation}.

Unlike machines, humans do an outstanding job of ignoring interfering signals and focusing on what we want to hear \cite{Waibel2009, Waibel2009a}. As deep learning gains popularity in speech signal processing, much impressive progress has been proposed to help enhance speech signals \cite{9189820} like denoising, dereverberation, source separation, and neural beamforming. Among speech enhancement techniques, masked-based is among the most popular and effective. In masking approaches, rather than estimating the enhanced signal directly, we estimate a mask, then multiply it with the noisy signal to get the enhanced signal. Depending on the type of input/output we can have waveform masking and spectral masking. Our study based on spectral masking since it's much faster than waveform masking.



Speech enhancement techniques generally use for blind signal enhancement. In our case study, we want to develop a robot to communicate and take orders from its master. So, we know precisely to whom the robot needs to listen, which can provide critical information that helps our speech recognition system work better, especially in complex acoustic situations like cocktail parties. 

There are some related studies for this circumstance, known as the \textit{speaker extraction} problem, including DENet \cite{Wang2018DeepEN}, SpeakerBeam \cite{8462661}, and VoiceFilter \cite{Wang2019, wang20z_interspeech}. However, our proposal has a few significant distinctions from them:  (1) we utilize an x-vector pre-trained model \cite{10.1007/978-3-030-59430-5_11} instead of i-vector or d-vector in \cite{Wang2018DeepEN, 8462661, Wang2019, wang20z_interspeech} since the x-vector brings better results in our experiment. (2) We use  scale-invariant source-to-noise
ratio (SI-SNR) \cite{Luo2018TaSNetTA} as a loss function because it is a speech enhancement evaluation metric and a training target that makes optimizing and choosing the best model more precise. (3) Different from \cite{Wang2018DeepEN, 8462661}, we focus on improving WER like \cite{Wang2019, wang20z_interspeech} but joint tuning ASR and speaker extraction model rather than optimizing the loss function. (4) We make a pre-trained self-supervised model based on wav2vec2 \cite{DBLP:conf/nips/BaevskiZMA20} architecture that works better for the noise acoustic condition. (5) For the mask estimation model, we use Conformer block \cite{gulati20_interspeech} rather than LSTM and CNN in \cite{Wang2019, wang20z_interspeech}. Furthermore, we introduce a cross-extraction mechanism between the reference signal and noisy signal for speaker embedding, which enhances the performance of our model, as demonstrated in our experiments.



\begin{figure*}[htpb]
  \centering
  \includegraphics[width=0.75\linewidth]{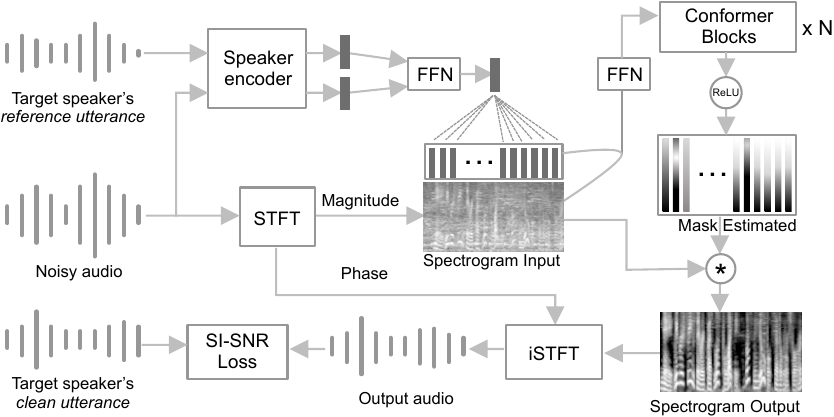}
\vspace{-3mm}
    \caption{Overview of the ConVoiFilter model.}
  \label{fig:model_overview}
\vspace{-4mm}
\end{figure*}

\vspace{-2mm}
\section{Model description}
\vspace{-3mm}

Our system consists of two primary modules: speaker extraction, which enhances the target speaker's voice, and the ASR module. In the following subsections, we will describe each module and our approach to jointly tuning them.

\vspace{-2mm}
\subsection{Target speaker's voice enhancement}
\vspace{-2mm}

Figure \ref{fig:model_overview} shows an overview of our target speaker's extraction module, ConVoiFilter. This module aims to remove all noise and interfering speech from the noisy audio input, producing a clean utterance for the target speaker. 

Firstly, an embedding vector identifies the target speaker ($e_{ref}\in\mathbb{R}^{d\_emb}$) extracted from their audio recordings (reference utterance) using a speaker encoder module. Moreover, we perform cross-extraction of speaker embedding from the noisy audio ($e_{noisy}\in\mathbb{R}^{d\_emb}$). Afterwards, we concatenate the two embeddings (from the reference utterance and the noisy audio) and feed them through a feed-forward network to produce the final presentation for the target speaker ($e = FFN(e_{ref}, e_{noisy})\in\mathbb{R}^{d\_emb}$). Without cross-extraction, $e = e_{ref}$. We experimented with two different speaker encoder models: x-vector \cite{10.1007/978-3-030-59430-5_11} and i-vector \cite{wan2018generalized}.

Secondly, we utilize the short-time Fourier transform (STFT) to process the noisy audio and generate a magnitude spectrogram ($ F\in\mathbb{R}^{S*(\frac{n\_fft}{2}+1)}$) that will aid in the mask estimation process. In this step, we preserve the phase ($P$) for the inverse STFT (iSTFT), which is used to reconstruct the output audio. The variable $S$ denotes the number of time frames, and $n\_fft$ represents the size of the Fourier transform. To create the features required to guide the mask estimation model, we horizontally concatenate the target speaker's embedding $E$ with the magnitude spectrogram $F$, resulting in $\hat{G} = \begin{bmatrix}F & e\end{bmatrix} \in \mathbb{R}^{S*(\frac{n\_fft}{2}+1+ d\_emb)}$. 





Next, $N$ conformer blocks used to transforms $\hat{G}$ into a mask $M\in\mathbb{R}^{S*(\frac{n\_fft}{2}+1)}$ of the same shape as the magnitude spectrogram $F$. Since $\hat{G}$ and $M$ is in different shape and we want to maintain the original conformer blocks, we need to transform $\hat{G}$ into $G = FFN(\hat{G}) \in\mathbb{R}^{S*(\frac{n\_fft}{2}+1)}$ which has the same shape with $M$. A conformer block (formula \ref{model:conformer:equation}) consists of four modules stacked together (described in \cite{gulati20_interspeech}) where $x_i$ is the input to conformer block $i$ ($x_0 = G$). We selected conformer because it incorporates convolution and multi-head self-attention, both of which are effective in utilizing contextual information, which is crucial for detecting interfering signals. The mask's purpose is to amplify or attenuate the amplitude of certain frequencies in the spectrogram, and therefore, it must be greater than or equal to zero. The ReLU function is applied to the output of the conformer block ($M = ReLU(y_N)$).

\vspace{-4mm}
\begin{equation}
\begin{split}
    \tilde{x_{i}}&= x_{i} + \frac{1}{2}\mathrm{FFN}(x_{i})    \\
    x'_{i}&= \tilde{x_{i}} + \mathrm{MHSA}(\tilde{x_{i}})    \\
    x''_{i}&= x'_{i} + \mathrm{Conv}(x'_{i})    \\
    y_{i}&= \mathrm{Layernorm}(x''_{i} + \frac{1}{2}\mathrm{FFN}(x''_{i}))    \\
\end{split}
\label{model:conformer:equation}
\end{equation}

Finally, the estimated mask $M$ is multiplied element-wise with the magnitude spectrogram $F$ to obtain an enhanced magnitude spectrogram. Phase information $P$ is combined with this enhanced magnitude spectrogram to reconstruct the output audio $iSTFT(M\odot F, P)$. We evaluate audio quality using the SI-SNR \cite{Luo2018TaSNetTA} loss function, which compares it with the clean utterance of the target speaker.

\vspace{-2mm}
\subsection{Automatic speech recognition}

Self-supervised learning of speech representations \cite{9893562} has recently shown its effectiveness in utilizing unlabeled speech data, resulting in outperforming the state-of-the-art (SoTA) in many automatic speech recognition (ASR) datasets. For our study, we utilized the pre-trained wav2vec2 model \cite{DBLP:conf/nips/BaevskiZMA20} to construct our ASR model. The wav2vec2 model acts as a speech encoder, and for the decoder, we used an RNN transducer \cite{https://doi.org/10.48550/arxiv.1211.3711}. Despite having a speech enhancement module to eliminate noise from the audio, the output may still contain noise. To address this issue, we utilized the self-supervised learning capabilities of wav2vec2 and created a pre-trained model by incorporating noise and room reverb into the unlabeled data (see section \ref{data} for dataset details). Our subsequent experiment demonstrated that this approach significantly enhances the system's accuracy.

\vspace{-2mm}
\subsection{Joint fine-tuning strategy}

ConVoiFilter is expected to produce only the target speaker’s voice. However, in practice, speech enhancement always comes up with unknown artifacts. In a naive way, we can directly connect the enhancing module to the ASR module and optimize their total loss. However, there are two reasons why it does not work. Firstly, the enhancement works in a high resolution of the input (e.g., an audio 15s, rate of 16kHz, STFT has hop size of 128 will output around 2000 time-frames). It makes the mask estimation module hard to work. Whereas with the wav2vec model, the same audio will output about 700 time-frames. Secondly, ConVoiFilter quickly fails the ASR module at the beginning of optimization because its output is too noisy. 

We handle these issues with a chunk-merging strategy. Below is the pseudo-code of the forward function. First, long audio is split into smaller chunks (line 4) to optimize the enhancing module, then the output (line 6) to optimize the ASR module. Audio input is padded into integer times the chunk's size to ensure all chunks are the same after splitting.
\begin{lstlisting}[language=Python]
def forward(noisy_audio, spk_embed,
            clean_audio, label):
  chunk_size = 5 #5s each chunk
  noisy_chunks = split(noisy_audio, chunk_size)
  output_chunks = enhance(noisy_chunks, spk_embed)
  output_audio = merge(output_chunks)
  enh_loss = si_snr(output_audio, clean_audio)
  if snr_loss < threshold: #enhancing is well
    output_text = asr(output_audio)
  else:
    output_text = asr(clean_audio)
  asr_loss = transducer_loss(output_text, label)
  loss = enh_loss + asr_loss
  return loss
\end{lstlisting}
In the beginning, the enhancement module can output randomly. So, we use a threshold to decide if the enhancement module works well; then, the ASR module will learn from its output; else, the ASR model will learn from the clean audio.

\vspace{-3mm}
\section{Experimental setup}
\vspace{-2mm}
\subsection{Data preparation}
\vspace{-1mm}
\label{data}
In our setup, we need a clean utterance of a specific speaker, noisy audio containing that utterance, and that speaker's embedding. Although CHiME-5 \cite{DBLP:journals/corr/abs-1803-10609} is like a cocktail party dataset; however, the clean utterance is no warranty. It shows through the WERs for the development set using the binaural microphones (clean utterance of a speaker) reported around 47.9\%. Instead, we train and evaluate the system using our generated data. We use audio from the LibriSpeech dataset \cite{7178964} (2338 speakers for training, 73 speakers for testing). The ambient noise dataset includes MUSAN and WHAM \cite{musan2015, DBLP:journals/corr/abs-1907-01160} (a total of 189 hours including music, speech, and environmental noise, 169 hours for training, 20 hours for testing). The reverb dataset is from Room RIR and BUT Speech@FIT \cite{7953152, 8717722} (2650 room impulse response signals, 2350 signals for training, 300 signals for testing). 

\begin{figure}[htpb]
  \includegraphics[width=0.9\linewidth]{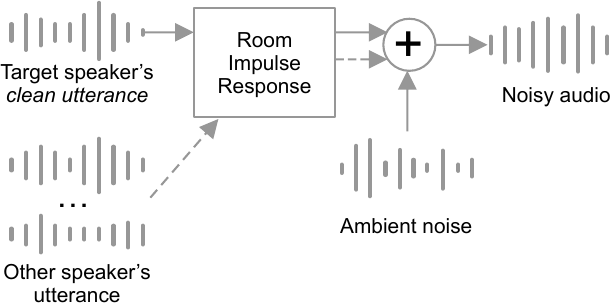}
  \vspace{-1mm}
    \caption{Data preparation pipeline}
  \label{fig:data_preparation}
\vspace{-2mm}
\end{figure}

Figure \ref{fig:data_preparation} shows our data generation pipeline. Noisy audio is cumulative of a few utterances and ambient noise. An utterance can be convolved with a room impulse response. We use random variables in the generating process to make the dataset more diverse. The number of other speakers' utterances is random, from 0 to 3. Ambient noise is added in 80\% number of times. The room impulse response is applied 30\% number of times. All other speaker’s utterances and ambient sound is normalized (based on the target speaker’s clean utterance) with SNR value around 1 to 20dB.

Both training processes (self-supervised wav2vec2 and ConVoiFilter model) use this data pipeline. One difference is that when we train self-supervised wav2vec2, we do not use other speakers' utterances (the dashed arrow in figure \ref{fig:data_preparation} - means no cross-talk) because it makes the data too noisy and can fail the wav2vec2 model.

\vspace{-2mm}
\subsection{Model setup}
\vspace{-2mm}
In our experiment, the ConVoiFilter model has 4 layer conformers with a hidden size is 1024, STFT has $n\_{fft}=512$, hope size is 128, speaker embedding is x-vector\cite{10.1007/978-3-030-59430-5_11} model has $d\_{emb}=512$. A speaker's embedding is calculated from multiple utterances of that speaker. For the ASR model, we use the wav2vec2-base architecture with 12 hidden layers; the hidden size is 768.

\vspace{-2mm}
\subsection{Evaluation}
\vspace{-2mm}
Our system was evaluated using five different types of data. The first is the ``Clean Audio'' which consists of the original audio. The second type is the ``Noisy Audio'' which is the output of the data processing pipeline (depicted in Figure \ref{fig:data_preparation}) applied to the clean audio. The third type is the ``No cross-talk'', a subset of noisy audio, which contains only single speaker samples with both ambient noise and speech reverberation. The fourth type is the ``Ambient noise'', a subset of noisy audio, which contains samples with cross-talk and ambient noise only. Finally, the fifth type is the ``Reverberation'', a subset of noisy audio, which contains samples with cross-talk and reverberation only.

\begin{table*}[]
\centering
\begin{tabular}{@{}|c|c|c|cc|c|@{}}
\toprule
\multirow{2}{*}{Model setting} & \multirow{2}{*}{Clean Audio} &
\multirow{2}{*}{No cross-talk} &
\multicolumn{2}{c|}{Cross-talk}                    & \multirow{2}{*}{Noisy Audio} \\ \cmidrule(lr){4-5}
                       &            &                  & \multicolumn{1}{c|}{Ambient noise} & Reverberation &                              \\ \midrule
ASR\_based     & 2.22  & 16.21                           & \multicolumn{1}{c|}{50.72}             & 90.87             & 80.04                            \\ \midrule
ASR\_noisy     & \textbf{2.03}  & 12.12                            & \multicolumn{1}{c|}{45.12}             & 84.14             & 75.19                            \\ \midrule
Cascade ConVoiFilter + ASR\_noisy        & 3.51    & 11.59                          & \multicolumn{1}{c|}{20.24}             & 30.30             & 26.40                            \\ \midrule
End-to-end ConVoiFilter-ASR\_noisy        & 3.36     & \textbf{9.41}                        & \multicolumn{1}{c|}{\textbf{13.23}}             & \textbf{25.14}             & \textbf{14.51}                            \\ \bottomrule
\end{tabular}
\vspace{-1mm}
\caption{\%WERs for different types of model settings.  The ConVoiFilter model uses x-vector as speaker encoder}
\label{tab:wer_result}
\vspace{-4mm}
\end{table*}

We evaluated our system using four different model settings to assess the WER on various types of data (table \ref{tab:wer_result}). The first two settings consisted of ASR models only, which aimed to measure the ASR model's ability to handle noisy data. The first ASR model, named ASR\_based, was initialized from the pre-trained wav2vec2 base model \cite{DBLP:conf/nips/BaevskiZMA20}, which was trained with 960 hours of Librispeech data. The second ASR model, ASR\_noisy, was initialized from our pre-trained wav2vec2 base model, which was trained with the same 960 hours of data, but augmented with noise and reverb data. The remaining two settings incorporated a speech enhancement module. The third was a cascade model, in which ConVoiFilter and ASR were trained independently. The final model was end-to-end, where ConVoiFilter and ASR were jointly trained. The first two ASR models were trained with noisy audio (without cross-talk). In contrast, the cascade and end-to-end models were trained with noisy audio that may have cross-talk.

\vspace{-3mm}
\section{Results}
\vspace{-3mm}

\begin{table}[]
\begin{tabular}{@{}|l|cccccc|@{}}
\toprule
\multicolumn{1}{|c|}{\multirow{2}{*}{System}} & \multicolumn{6}{c|}{Overlap ratio in \%}   \\ \cmidrule(l){2-7} 
\multicolumn{1}{|c|}{}                        & 0S   & 0L   & 10   & 20    & 30    & 40    \\ \midrule
Baseline \cite{9053426}                                      & 8.4  & 8.3  & 11.6 & 15.8  & 18.7  & 21.7  \\ \midrule
Whisper-large                                       & \textbf{3.64} & \textbf{3.4}  & 8.86 & 15.64 & 23.55 & 32.73 \\ \midrule
\begin{tabular}[c]{@{}l@{}}ConVoiFilter +\\ Whipser-large\end{tabular} & 5.35 & 5.59 & \textbf{7.32} & \textbf{13.28} & \textbf{14.46} & \textbf{16.83} \\ \bottomrule
\end{tabular}
\caption{\%WERs for LibriCSS utterance-wise evaluation}
\label{tab:libricss}
\vspace{-5mm}
\end{table}

The WER for different model settings on various data types is presented in Table \ref{tab:wer_result}. Generally, the end-to-end models perform better than other models across most input sets, except for clean audio. In cases with cross-talk, the ConVoiFilter model demonstrates its effectiveness by using speaker embedding to extract the target speaker's voice, resulting in cleaner audio and a significant improvement in WER (the result in the cross-talk column, rows 2 and 4). The ``Noisy Audio'' column displays the overall model performance, with only the ASR model achieving the best WER at 75.19\% (ASR\_noisy). When combined with ConVoiFilter model, the WER can be significantly reduced to 26.40\%. The end-to-end model can further improve the performance, achieving a WER of 14.51\%. Additionally, our pre-trained wav2vec2 model (with noisy audio) proves its worth, with ASR\_noisy outperforming ASR\_based in all input sets.

Table \ref{tab:libricss} presents the WER for the LibriCSS\cite{9053426} dataset, a 10-hour real-recorded dataset derived from the LibriSpeech corpus featuring speaker conversations. The baseline comprises BLSTM ASR, speaker separation, and MVDR based on 7-channels. We compare this baseline with a robust ASR model (Whisper large \cite{radford2022robust}) and also explore a combination of our ConVoiFilter with Whisper. The results clearly indicate that ConVoiFilter significantly improves Whisper, particularly in reducing WER in overlapping audio.

Table \ref{tab:enh_result} presents an ablation study comparing our proposal to other studies on the speech enhancement model for a target speaker (speaker extraction problem). The effectiveness is measured using two standard metrics, SI-SNR and SDR, expressed in dB, where a higher value indicates better performance. Our ConVoiFilter differs from the recent VoiceFilter \cite{Wang2019, wang20z_interspeech} in three key aspects, namely the speaker encoder (x-vector), mask estimation model (conformer), and loss function (SI-SNR loss). Table \ref{tab:enh_result} demonstrates the improvement resulting from each change we made. Firstly, replacing bi-LSTM with Conformer resulted in the most significant gain (5.56 points in SDR). Secondly, the SI-SNR loss function outperformed the MSE loss used in \cite{Wang2019}. We speculate that the direct computation of the audio signal by the SI-SNR loss provides better optimization signals than the spectrogram-based MSE loss. Thirdly, although the x-vector and i-vector were trained with the same dataset, our experiment showed that the x-vector provided better results than the i-vector.

\begin{figure}[htpb]
  \includegraphics[width=1.0\linewidth]{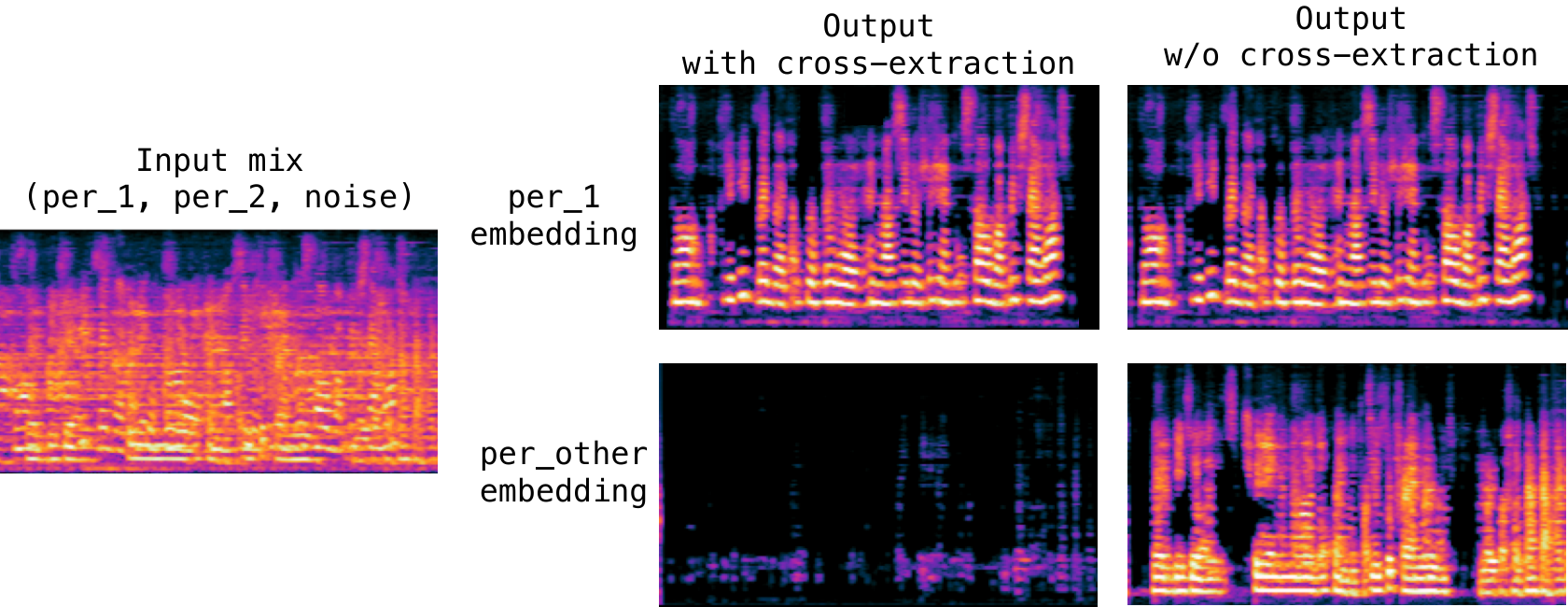}
  \vspace{-6mm}
    \caption{An illustration spectrogram demonstrating the extraction of the target speaker's voice.}
  \label{fig:example}
\vspace{-6mm}
\end{figure}

Figure \ref{fig:example} illustrates the effectiveness of cross-extraction of speaker embedding. In this example, the input is a mixture of two people (labeled as per\_1 and per\_2) and ambient noise. When we use the speaker embedding from per\_1, both the model with and without cross-extraction can extract the target speaker as per\_1. However, if the speaker embedding is obtained from a random person, only the model with cross-extraction can output a blank speech signal. Without the cross-extraction mechanism, the model extracts the wrong speech signal (the speech signal of per\_2).

\begin{table}[]
\centering
\begin{tabular}{@{}|c|c|c|@{}}
\toprule
Method          & SI-SNR & SDR \\ \midrule
No Enhancement  & 1.04      & 1.14   \\ \midrule
ConVoiFilter & \textbf{13.97}      & \textbf{15.14}   \\ \midrule
\makecell{x-vector $\rightarrow$ i-vector}        & 10.02      & 11.14   \\ \midrule
Conformer $\rightarrow$ bi-LSTM            & 8.12      & 9.58   \\ \midrule
SI-SNR loss $\rightarrow$ MSE loss        & 9.11      & 10.81   \\ \bottomrule
\end{tabular}
\vspace{1mm}
\caption{Ablation study over each change in the ConVoiFilter model.  Source to distortion ratio (SDR) and Scale invariant signal to noise ratio
(SI-SNR) in dB.}
\label{tab:enh_result}
\vspace{-5mm}
\end{table}

\section{Conclusion}
\vspace{-3mm}
This paper details a case study on cocktail party speech recognition. Instead of recognizing all speakers, our system focuses on enhancing the target speaker's voice before conducting speech recognition. Through rigorous experiments, we showcased the effectiveness of our improved end-to-end model. Noteworthy enhancements include a cross-extraction speaker encoder, an improved mask estimation model, and an optimized loss function. We also publicly share our pre-trained ConVoiFilter to support ongoing research.

\vspace{-3mm}
\section{ACKNOWLEDGMENT}
\vspace{-2mm}
The authors gratefully acknowledge the support provided by Carl Zeiss Stiftung under the project Jung bleiben mit Robotern (P2019-01-002).


\bibliographystyle{IEEEbib}
\bibliography{strings,refs}

\begin{thebibliography}{10}

\bibitem{DBLP:conf/interspeech/NguyenSW21}
Thai{-}Son Nguyen, Sebastian St{\"{u}}ker, and Alex Waibel,
\newblock ``Super-human performance in online low-latency recognition of conversational speech,''
\newblock in {\em Interspeech 2021}. 2021, {ISCA}.

\bibitem{yang1999multimodal}
Jie Yang, Alex Waibel, and et~al.,
\newblock ``Multimodal people id for a multimedia meeting browser,''
\newblock in {\em Proceedings of the seventh ACM international conference on Multimedia (Part 1)}, 1999, pp. 159--168.

\bibitem{bett2000multimodal}
Michael Bett, Alex Waibel, and et~al.,
\newblock ``Multimodal meeting tracker.,''
\newblock in {\em RIAO}. Citeseer, 2000, pp. 32--45.

\bibitem{zorilua2019investigation}
C{\u{a}}t{\u{a}}lin Zoril{\u{a}}, Christoph Boeddeker, Rama Doddipatla, and Reinhold Haeb-Umbach,
\newblock ``An investigation into the effectiveness of enhancement in asr training and test for chime-5 dinner party transcription,''
\newblock in {\em 2019 IEEE (ASRU)}. IEEE, 2019.

\bibitem{Waibel2009}
Alex Waibel, Hartwig Steusloff, Rainer Stiefelhagen, and Kym Watson,
\newblock {\em Computers in the Human Interaction Loop}, pp. 3--6,
\newblock Springer London, London, 2009.

\bibitem{Waibel2009a}
Alex Waibel,
\newblock {\em Beyond CHIL}, pp. 367--371,
\newblock Springer London, London, 2009.

\bibitem{9189820}
Reinhold Haeb-Umbach, Jahn Heymann, Lukas Drude, Shinji Watanabe, Marc Delcroix, and Tomohiro Nakatani,
\newblock ``Far-field automatic speech recognition,''
\newblock {\em Proceedings of the IEEE}, vol. 109, no. 2, pp. 124--148, 2021.

\bibitem{Wang2018DeepEN}
J.~Wang, Jie Chen, Dan Su, Lianwu Chen, Meng Yu, Yanmin Qian, and Dong Yu,
\newblock ``Deep extractor network for target speaker recovery from single channel speech mixtures,''
\newblock in {\em INTERSPEECH}, 2018.

\bibitem{8462661}
Marc Delcroix and et~al.,
\newblock ``Single channel target speaker extraction and recognition with speaker beam,''
\newblock in {\em 2018 IEEE International Conference on Acoustics, Speech and Signal Processing (ICASSP)}, 2018, pp. 5554--5558.

\bibitem{Wang2019}
Quan Wang and et~al.,
\newblock ``{VoiceFilter: Targeted Voice Separation by Speaker-Conditioned Spectrogram Masking},''
\newblock in {\em Proc. Interspeech 2019}, 2019, pp. 2728--2732.

\bibitem{wang20z_interspeech}
Quan Wang and et~al.,
\newblock ``{VoiceFilter-Lite: Streaming Targeted Voice Separation for On-Device Speech Recognition},''
\newblock in {\em Proc. Interspeech 2020}, 2020, pp. 2677--2681.

\bibitem{10.1007/978-3-030-59430-5_11}
Juan~M. Coria and et~al.,
\newblock ``A comparison of metric learning loss functions for end-to-end speaker verification,''
\newblock in {\em Statistical Language and Speech Processing}, Luis Espinosa-Anke, Carlos Mart{\'i}n-Vide, and Irena Spasi{\'{c}}, Eds., Cham, 2020, pp. 137--148, Springer International Publishing.

\bibitem{Luo2018TaSNetTA}
Yi~Luo and Nima Mesgarani,
\newblock ``Tasnet: Time-domain audio separation network for real-time, single-channel speech separation,''
\newblock {\em 2018 IEEE International Conference on Acoustics, Speech and Signal Processing (ICASSP)}, pp. 696--700, 2018.

\bibitem{DBLP:conf/nips/BaevskiZMA20}
Alexei Baevski, Yuhao Zhou, Abdelrahman Mohamed, and Michael Auli,
\newblock ``wav2vec 2.0: {A} framework for self-supervised learning of speech representations,''
\newblock in {\em NeurIPS 2020}, 2020.

\bibitem{gulati20_interspeech}
Anmol Gulati and et~al.,
\newblock ``{Conformer: Convolution-augmented Transformer for Speech Recognition},''
\newblock in {\em Proc. Interspeech 2020}, 2020, pp. 5036--5040.

\bibitem{wan2018generalized}
Li~Wan, Quan Wang, Alan Papir, and Ignacio~Lopez Moreno,
\newblock ``Generalized end-to-end loss for speaker verification,''
\newblock in {\em 2018 IEEE International Conference on Acoustics, Speech and Signal Processing (ICASSP)}. IEEE, 2018, pp. 4879--4883.

\bibitem{9893562}
Abdelrahman Mohamed and et~al.,
\newblock ``Self-supervised speech representation learning: A review,''
\newblock {\em IEEE Journal of Selected Topics in Signal Processing}, pp. 1--34, 2022.

\bibitem{https://doi.org/10.48550/arxiv.1211.3711}
Alex Graves,
\newblock ``Sequence transduction with recurrent neural networks,'' 2012.

\bibitem{DBLP:journals/corr/abs-1803-10609}
Jon Barker, Shinji Watanabe, Emmanuel Vincent, and Jan Trmal,
\newblock ``The fifth 'chime' speech separation and recognition challenge: Dataset, task and baselines,''
\newblock {\em CoRR}, vol. abs/1803.10609, 2018.

\bibitem{7178964}
Vassil Panayotov, Guoguo Chen, Daniel Povey, and Sanjeev Khudanpur,
\newblock ``Librispeech: An asr corpus based on public domain audio books,''
\newblock in {\em 2015 IEEE International Conference on Acoustics, Speech and Signal Processing (ICASSP)}, 2015, pp. 5206--5210.

\bibitem{musan2015}
David Snyder, Guoguo Chen, and Daniel Povey,
\newblock ``{MUSAN}: {A} {M}usic, {S}peech, and {N}oise {C}orpus,'' 2015,
\newblock arXiv:1510.08484v1.

\bibitem{DBLP:journals/corr/abs-1907-01160}
Gordon Wichern and et~al.,
\newblock ``Wham!: Extending speech separation to noisy environments,''
\newblock {\em CoRR}, vol. abs/1907.01160, 2019.

\bibitem{7953152}
Tom Ko and et~al.,
\newblock ``A study on data augmentation of reverberant speech for robust speech recognition,''
\newblock in {\em 2017 IEEE International Conference on Acoustics, Speech and Signal Processing (ICASSP)}, 2017, pp. 5220--5224.

\bibitem{8717722}
Igor Szöke, Miroslav Skácel, Ladislav Mošner, Jakub Paliesek, and Jan Černocký,
\newblock ``Building and evaluation of a real room impulse response dataset,''
\newblock {\em IEEE Journal of Selected Topics in Signal Processing}, vol. 13, no. 4, pp. 863--876, 2019.

\bibitem{9053426}
Zhuo Chen and et~al.,
\newblock ``Continuous speech separation: Dataset and analysis,''
\newblock in {\em ICASSP 2020}, 2020, pp. 7284--7288.

\bibitem{radford2022robust}
Alec Radford, Jong~Wook Kim, Tao Xu, Greg Brockman, Christine McLeavey, and Ilya Sutskever,
\newblock ``Robust speech recognition via large-scale weak supervision,'' 2022.

\end{thebibliography}

\end{document}